\documentclass[twocolumn,showpacs,preprintnumbers,amsmath,amssymb]{revtex4}

\usepackage{graphicx}
\usepackage{dcolumn}
\usepackage{bm}

\begin{document}
\preprint{APS/123-QED}

\title{Deconstruction of the Kondo Effect near the AFM-Quantum Critical Point}

\author{H. Maebashi,$^{1}$ K. Miyake,$^2$ and C.M. Varma$^{3}$}

\affiliation{%
$^1$Institute for Solid State Physics, University of Tokyo, 
Kashiwa, Chiba 277-8581, Japan
\\
$^2$Department of Physical Science, Graduate School of Engineering Science,
Osaka University, Toyonaka, Osaka 560-8531, Japan
\\
$^3$Department of Physics, University of California, 
Riverside, California 92521
}%

\date{\today}

\begin{abstract}
The problem of a spin-1/2 magnetic impurity  near an antiferromagnetic transition of the
host lattice is shown to transform to a multichannel problem. A variety of fixed points is discovered asymptotically near the AFM-critical point. Among these is a new variety of stable fixed point of  a multichannel Kondo problem which does not require channel isotropy. At this point Kondo screening disappears but coupling to spin-fluctuations remains.
Besides its intrinsic interest, the problem is an essential ingredient in the problem of quantum critical points in heavy-fermions.
\end{abstract}

\maketitle

Theories of quantum critical phenomena involving fermions rely on extensions of the theory of
classical critical phenomena to quantum problems  by integrating out the fermions in favor of low energy bosonic fluctuations \cite{Hertz76, Moriya85}. 
Several experimental results on heavy-fermion quantum critical points (QCP's) are in disagreement with such theories \cite{expts, VNV02}.  
Some valiant efforts to address the problem are being made by using an extended dynamical mean field theory \cite{Si01}. 
As an interesting problem in itself as well as to gain insight to the difficult problem of the lattice, we present here a systematic theory 
for a single impurity in a host with a diverging antiferromagnetic (AFM) correlation length. 
We show below that this necessarily leads to a multi-channel problem with a variety of remarkable properties.

The $S=1/2$ localized moment is coupled to the host itinerant electrons, which are near an AFM instability due to electron-electron interactions, 
by the Hamiltonian
$(J/2)(2 \pi)^{-d} \int{\rm d}^d k (2 \pi)^{-d}\int{\rm d}^d k'
\psi_{{\bf k}}^\dagger {\mbox{\boldmath $\sigma$}}\psi_{{\bf k}'}^{\phantom \dagger}
\cdot {\bf S}$  
where $\psi_{{\bf k}}^\dagger$ ($\psi_{{\bf k}'}^{\phantom \dagger}$) is 
the creation (annihilation) operator for the itinerant electron with momentum 
${\bf k}$ (${\bf k}'$), and 
${\bf S}$ is the localized spin. Due to the interactions among the host electrons, 
the bare Kondo vertices $J$ are renormalized. 
For a Fermi liquid such vertex corrections lead only to a numerical renormalization. 
However, qualitatively new effects can arise due to such renormalizations 
in the vicinity of a QCP in the host itinerant electrons. 
This problem has been solved for the case of a ferromagnetic instability of the itinerant electrons \cite{LM72,MMV02}. 
The problem of the AFM instability is both physically and technically quite different.
 
In order to perform a renormalization group (RG) procedure, 
we first derive the low-energy effective action in which the momenta of the itinerant electrons 
are restricted within a narrow region near the Fermi surface, $|\varepsilon_{\bf k}| \leq W$ ($\varepsilon_{\bf k}$ 
is the energy of the host electrons relative to the Fermi level). 
Denote $\psi_{\bf k}$ ($\psi_{\bf k}^{\dagger}$) as $\psi_{<}$ ($\psi_{<}^{\dagger}$) for $|\varepsilon_{\bf k}| \leq W$ 
otherwise $\psi_{>}$ ($\psi_{>}^{\dagger}$), then the Hamiltonian can be expressed as 
$H [\psi^{\phantom \dagger}, \psi^{\dagger}] = H [\psi_{<}^{\phantom \dagger}, \psi_{<}^{\dagger}] 
+ H [\psi_{>}^{\phantom \dagger}, \psi_{>}^{\dagger}] 
+ H' [\psi_{<}^{\phantom \dagger}, \psi_{<}^{\dagger}, \psi_{>}^{\phantom \dagger}, \psi_{>}^{\dagger}]$.
Eliminating the modes for $|\varepsilon_{\bf k}| > W$ in a path-integral formulation, we obtain the effective action 
as ${\cal A}_{eff} = {\cal A}_{eff}^{(0)} + {\cal A}_{eff}^{(cc)} + {\cal A}_{eff}^{(cf)} + .......$; 
${\cal A}_{eff}^{(0)}$ describes the free parts of the effective action, 
${\cal A}_{eff}^{(cc)}$ is the mutual-interaction term among the spins of the host itinerant electrons 
and ${\cal A}_{eff}^{(cf)}$ corresponds to the effective interaction between the host electrons and the localized spin 
which is represented by Fig.\ref{fig:1}:

\begin{figure}[b]
\includegraphics[width=7.5cm]{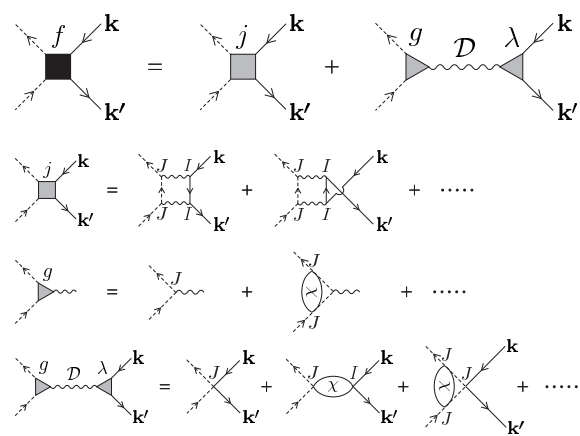}
\caption{\label{fig:1}Exchange interaction between the host electrons and the localized spin in the low-energy effective action. 
The first line gives the complete interaction; the subsequent lines show first few terms of the series that is summed in each vertex 
in the first line. $\chi$ represents the dynamical spin susceptibility of the host electrons.}
\end{figure}

\begin{figure*}
\includegraphics[width=16.0cm]{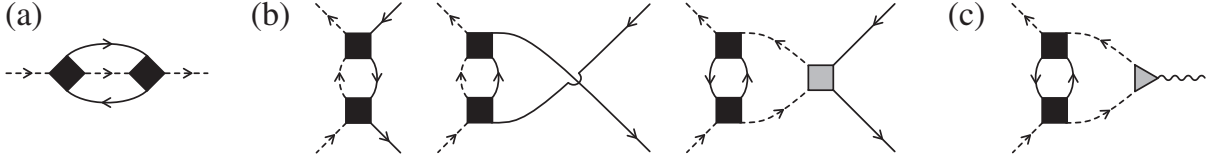}
\caption{\label{fig:2}Perturbative corrections to (a) the self-energy of the local moment represented in terms of pseudo fermions,  
(b) the vertex $j$, and (c) the vertex $g$ describing the coupling of 
the local moment to the magnetic fluctuations.}
\end{figure*}

Fig.\ref{fig:1} presents Feynman diagrams for $A_{eff}^{(cf)}$ in which 
solid and broken lines are associated with the host electron and the local moment respectively. 
The exchange couplings in $A_{eff}^{(cf)}$ has two parts: 
the first term $j$ in Fig.\ref{fig:1} is irreducible with respect to the  propagator ${\cal D}_{{\bf k},{\bf k}'}(\omega)$ describing 
the AFM spin fluctuations of the host electrons, while the second part is reducible. 
The division into the two parts is such that $g$ starts out as $J$  while $j$ starts out as $O(J^2)$, 
as explained through the first few terms of the series for $j$, and $g$ in the figure. 
$g$ is the coupling of the spin-fluctuations to the local moments which are always coupled through electron-electron interaction
vertices $\lambda$ to the fermions. 
The vertices $g$, and $\lambda$ are also irreducible 
with respect to the propagator ${\cal D}_{{\bf k},{\bf k}'}$. 
  $\lambda$ = 
$I \Lambda^{>}_{\phantom K}(k, k')$ where $I$ is the bare interaction among the spins of the host electrons 
and $\Lambda^{>}_{\phantom K}(k, k')$ is 
the one-interaction irreducible vertex part in the spin channel for $H [\psi_{>}^{\phantom \dagger}, \psi_{>}^{\dagger}]$.
${\cal D}_{{\bf k},{\bf k}'}(\omega)$ is related to the dynamical spin susceptibility 
$\chi_{{\bf k},{\bf k}'}^{>}(\omega)$ 
of the host described by $H [\psi_{>}^{\phantom \dagger}, \psi_{>}^{\dagger}]$ as
\begin{equation} 
{\cal D}_{{\bf k},{\bf k}'}(\omega) = [ 1 + I \chi_{{\bf k},{\bf k}'}^{>}(\omega)]/I\;.
\end{equation}  
Note that even in the limit of $W \to 0$ near the QCP, $\Lambda^{>}_{\phantom K}$ is 
not singular (while $\chi^{>}$ is).
So, we can safely neglect the dependence of $\lambda$ on the cutoff as well as the momenta, 
after ${\cal D}_{{\bf k},{\bf k}'}$ is extracted.  

In the limit that the energy cutoff $W \to 0$,  the momenta ${\bf k}$ and ${\bf k}'$ are restricted to be on the Fermi surface $S_{\rm F}$. 
If ${\bf k}$ is represented by its projection onto $S_{\rm F}$ denoted by ${\bf K}$ and 
the energy shell it belongs to $\varepsilon = \varepsilon_{\bf k}$, 
we can approximate ${\cal D}_{{\bf k}, {\bf k}'} (\omega) \sim {\cal D}_{{\bf k}, {\bf k}'} (0) \equiv {\cal D}({\bf K}, {\bf K}')$ as  
\begin{equation}
{\cal D}({{\bf K}, {\bf K}'}) = 
\frac{A N_0}{\kappa(W)^2 + 2 \left[ d +  \sum_{i=1}^{d} \cos (K_i^{\phantom *} - K'_i) \right]},
\label{spectra}
\end{equation} 
where $\kappa(W)$ is the inverse magnetic correlation length,  $N_0$ is the density of states, $A$ is a constant of the order of $1$. 
The retardation of the interactions is properly included through a cut-off which appears through $\kappa(W) \propto W^z$, 
where $z$ is the dynamical exponent.
This procedure has been explicitly justified in Ref.\cite{MMV02} where it occurs in the same context. 
 
Now, consider the unitary transformation which diagonalizes 
${\cal D}({\bf K},{\bf K}') $ on the Fermi surface $S_{\rm F}$.
Assuming that the impurity sits in a site with the full point group symmetry of the lattice, 
the symmetry operations $R$ of a point group $G$ (say, $C_{4v}$ for the $d$ = $2$ square symmetry) may be used such that 
it is enough to find eigenvalues of ${\cal D}({\bf K},{\bf K}')$ with ${\bf K}$ restricted to an irreducible  portion of the Brillouin zone, $\Omega$ 
 (a triangle determined by 
the vertices $(0, 0)$, $(\pi, 0)$, and $(\pi, \pi)$ for $C_{4v}$).
We obtain
\begin{equation}
\int_{{\bf K}' \in \Omega} \sum_{m'=1}^{d_{\alpha}}
{\cal D}_{\alpha}^{\phantom *} ({\bf K}, {\bf K}')_{m, m'}^{\phantom *} 
u_{l}^{\phantom \dagger} ({\bf K}')_{m'}^{\phantom \dagger} 
= 
{\cal D}_{l}^{\phantom *}(\kappa) u_{l}^{\phantom \dagger} ({\bf K})_{m}^{\phantom \dagger} \;,
\label{eigen3}
\end{equation}
Here $\int_{{\bf K}}$ stands for the average over $S_{\rm F}$, $N_0^{-1}(2\pi)^{- d}{\int_{\bf k}}\delta(\varepsilon_{\bf k})$; 
${\cal D}_{\alpha}^{\phantom *} ({\bf K}, {\bf K}')_{m, m'}^{\phantom *} = 
\sum_{R \in G} R{\cal D}({\bf K}, {\bf K}') \Gamma_{\alpha}^{\phantom *} (R)_{m, m'}^{*}$. 
$\Gamma_\alpha^{\phantom \dagger} (R)$
is the unitary matrix for the irreducible representation $\alpha$, the dimension of which is $d_{\alpha}$. 
Therefore, $l$ can be represented by a set of $\alpha$ and $i$ in which 
$i$ tells apart eigenvalues in the space of $\{ 1, 2, \ldots, d_{\alpha} \}$ $\otimes$ $\{ {\bf K} | {\bf K} \in \Omega\}$ for each $\alpha$.  
In the whole space of the Fermi surface, 
the number of degeneracies $d_l$ is equal to $d_{\alpha}$. 
This general result is always true but more interesting is the generic case in which  the AFM vectors ${\bf Q}$, 
(assumed commensurate \cite{AIM95}) connects points on the Fermi-surface ("hot-spots" in 2-d but "hot-lines" in 3-d). 
In that case the problem acquires larger degeneracies. 
 
Expanding $\psi_{\bf k}$ as $\psi_{\bf k}$ = $\psi_{{\bf K}, \varepsilon}$ = 
$\sum_{l, m}u_{l}^{\phantom *}({\bf K})_{m}^{\phantom *} \psi_{l, m, \varepsilon}^{\phantom *}$, 
the equation represented by Fig.\ref{fig:1} leads to the effective interaction of the local moment:
\begin{equation} 
\frac{1}{2} \sum_{l,m}  
\left[
j_l + g \lambda {\cal D}_{l}(\kappa) 
\right]
a_{l,m}^{\dagger} {\mbox{\boldmath $\sigma$}} a_{l,m}^{\phantom \dagger}
\cdot  
{\bf S}\,,
\label{Heff}
\end{equation}
where $a_{l,m}^{\phantom \dagger}$
is defined by
$a_{l,m}^{\phantom \dagger}$ $\equiv$ $\int \psi_{l,m,\varepsilon} {\rm d} \varepsilon$. 
At the initial cutoff $W_0$, $g$ = $O(J)$ and $j_l$ = $O(J^2)$ for all $l$ so that $g$ $>>$ $j_l$ for weak couplings $J$.
Eq.(\ref{Heff}) has the form of a multichannel Kondo Hamiltonian with the number of channel $d_{l}$ 
for each $l$, but the interactions depend {\it explicitly} on $W$ through $\kappa$.  

The RG equations for Eq.(\ref{Heff}) are now derived for any given  ${\cal D}_{l}(\kappa) $.  
Fig.\ref{fig:2} presents perturbative corrections up to the third order of $j$ and $g$ in the successive 
elimination of modes for $W' < \varepsilon_{\bf k} \leq W$. Define a crossover parameter  $W_1$ $\sim$ $r W_0$; $r$ is the distance from the QCP.  
For the present case of $z=2$, $\kappa/\kappa_0$ $=$ $\sqrt{W/W_0}$ for $W > W_1$, i.e. 
the "quasi-classical" regime while $\kappa/\kappa_0 $ $=$ $\sqrt{r}$ 
for $W < W_1$, i.e. the quantum regime. $\kappa_0$ is a constant of the order of $1$. 
In this paper we present results only for the "quasi-classical" regime . It is also useful to introduce  
$\epsilon$ $\equiv$ $({\rm d}/{\rm d} t)$ $\ln \left[ \textstyle{\lambda^2 \sum_l d_l {\cal D}_l^2} \right]$ 
with $t = \ln (W_0 / W)$.
Because the imaginary part of the local spin susceptibility for the host electrons scales 
as ${\rm Im}{\chi}_{loc}(\omega) \sim \lambda^2 \sum_{l} d_{l}^{\phantom *} {\cal D}_l^2 \omega \sim W^{- \epsilon} \omega$, 
$\epsilon$  determines the power law of $\chi_{loc}(\tau)$ for the long-time limit $\tau \to \infty$
as $\chi_{loc}(\tau)$ $\sim$ $1/\tau^{(2 - \epsilon)}$.
Note that $\epsilon$ $=$ $(4 - d)/2$ for $W > W_1$,  
(while $\epsilon = 0$ for $W < W_1$ as  for Fermi liquids.)

It is convenient to write the RG equations in terms of ${\bar t} = \ln (\kappa_0/\kappa)^2$  and to rescale $g$ and ${\cal D}_l$ as 
${\bar g}$ = $g \lambda \sqrt{\sum_l d_l {\cal D}_l^2}$ and 
${\bar {\cal D}}_l$ = ${\cal D}_l / \sqrt{\sum_l d_l {\cal D}_l^2}$. From the definition of $\epsilon$, it follows that $\sum_l d_l {\cal D}_l^{2}$ = $\int_{{\bf K}, {\bf K}'}{\cal D}({\bf K}, {\bf K}')^2$ 
$\propto$ $1/\kappa^{2 \epsilon}$.

After some manipulations the RG equations are derived as
\begin{subequations}
\label{beta}
\begin{eqnarray}
\frac{ {\rm d} j_l }{ {\rm d} {\bar t} } &= & 
N_0 f_l^2 - \frac{1}{2} N_0^2 j_l \sum_{l'} d_{l'} f_{l'}^2,
\\
\frac{ {\rm d} {\bar g} }{ {\rm d} {\bar t} } &= &
{\bar g} \left[ \frac{\epsilon}{2} - \frac{1}{2} N_0^2 \sum_{l'} d_{l'} f_{l'}^2 \right],
\label{beta2}
\end{eqnarray}
\end{subequations}
where $f_l = j_l + {\bar g} {\bar {\cal D}}_l$. 

\begin{figure}[t]
\includegraphics[width=7.5cm]{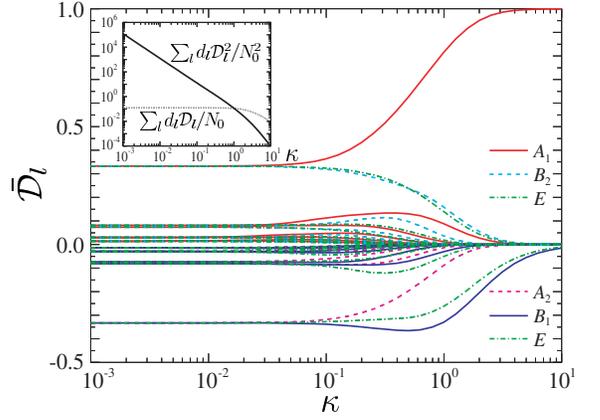}
\caption{\label{fig:3} ${\bar {\cal D}}_l$ = ${\cal D}_l / \sqrt{\sum_l d_l {\cal D}_l^2}$ versus $\kappa$ in a two-dimensional system 
with a circular Fermi line at half filling. 
The largest $7d_l$ eigenvalues for each symmetry are plotted.
The inset shows dependences of $\sum_l d_l {\cal D}_l^2/N_0^2$ and $\sum_l d_l {\cal D}_l/N_0$ on $\kappa$.}
\end{figure}

In order to solve Eqs.(\ref{beta}), we must first find ${\bar {\cal D}}_l(\kappa)$ from Eq.(\ref{eigen3}). 
We have done this for 2-d as well as 3-d analytically for
$\kappa \rightarrow 0$ and checked it numerically for a range of $\kappa$. Fig.\ref{fig:3} shows dependence of ${\bar {\cal D}}_l$ 
on $\kappa$ obtained from numerical solutions of Eq.(\ref{eigen3}) in a 2d square lattice with a circular Fermi line at half filling 
(the Fermi radius is $\sqrt{2\pi}$).  
In this case the Fermi surface has 8 "hot-spots", which are four pairs of points   connected by the AFM  wave-vector ${\bf Q}$. 
In the case of $C_{4v}$, 
there exist five irreducible representations in which
$d_{\alpha}$ = $1$ for $\alpha$ = $A_1$, $A_2$, $B_1$, and $B_2$ while  $d_{\alpha}$ = $2$ for $\alpha$ = $E$.  
We find quite clearly
that the absolute values of eight eigenvalues approach each other 
and $A_1$, $B_2$, and $E$ ($A_2$, $B_1$, and $E$) states bunch up to constitute four-fold degenerate states
with a positive (negative) value of ${\bar {\cal D}}_l$ in the limit of $\kappa \to 0$. 
From the inset it can be checked that the results are consistent with $\sum_l d_l {\cal D}_l^2$ $\propto$ $1/\kappa^2$, i.e., $\epsilon$ = $1$, 
so that $|{\cal D}_l|$ diverge as $1/\kappa$ for all $l$. 
 
The above realization of symmetry higher than that of the underlying lattice near the QCP 
can be understood from a general point of view; the 2-d example is explained here. Consider a Fermi line with $N_{\rm h}$ = $2n_{\rm h}$ 
equivalent hot spots, divided into $n_{\rm h}$ pairs with one member of the pair  connected to the other by ${\bf Q}$.
Let ${\bf K}_{\rm h}^{1}$ and ${\bf K}_{\rm h}^{2 }$ be the vectors of two hot spots 
of one such   pair, i.e., ${\bf K}_{\rm h}^{1}$ = ${\bf K}_{\rm h}^{2}$ + ${\bf Q}$, and 
${\bf e}_{\rm h}^{1}$ and ${\bf e}_{\rm h}^{2}$ be the unit vectors tangent to the Fermi line at these two hot spots. 
If we write ${\bf K}$ = ${\bf K}_{\rm h}^{\xi}$ + $p\,{\bf e}_{\rm h}^{\xi}$ and 
${\bf K}'$ = ${\bf K}_{\rm h}^{\eta}$ + $p'{\bf e}_{\rm h}^{\eta}$ in which 
$\xi$ = $1$ and $\eta$ = $2$ or vice-versa near these hot spots, 
then the singular parts of ${\cal D}({\bf K}, {\bf K}')$ can be approximated by
\begin{equation}
{\cal D}({\bf K}, {\bf K}')
\simeq \frac{N_0}{
\kappa^2 + p^2 + p^{\prime 2} - 
2 p^{\phantom \prime} p^{\prime} \cos \theta_{\rm h}
}, 
\label{approxchi}
\end{equation}
where 
$\cos \theta_{\rm h}$ is given by ${\bf e}_{\rm h}^{1} \cdot {\bf e}_{\rm h}^{2}$.
Substituting Eq.(\ref{approxchi}) into Eq.(\ref{eigen3}), 
and dividing it by $\sqrt{\sum_l d_l {\cal D}_l^2}$ $\propto$ $\kappa^{-1}$, 
the eigenvalues of ${\bar {\cal D}}_l(\kappa)$ in the limit of $\kappa \to 0$ can be obtained 
by solving the following equation:
\begin{equation}
\int_{-\infty}^{\infty} {\rm d} y \sum_{\eta = 1}^{2}
{\bar {\cal D}}(x,y)^{\xi}_{\eta}
u_l^{\phantom *}(y)_{m}^{\eta} 
=
{\bar {\cal D}}_{l}(0) 
u_l^{\phantom *}(x)_{m}^{\xi}, 
\label{eigenQCP}
\end{equation}
where $u_l^{\phantom *}(x)_m^{\xi} = u_l({\bf K}_{\rm h}^{\xi} + \kappa x {\bf e}_{\rm h}^{\xi})_m$;    
the kernel ${\bar {\cal D}}(x,y)^{\xi}_{\eta}$ is given by
\begin{equation}
{\bar {\cal D}}(x,y)^{\xi}_{\eta}
= \sqrt{ \frac{ \sin \theta _{\rm h} }{ 8 \pi } }
\frac{ 1 - \delta^{\xi}_{\eta} }{1 + x^2 + y^2 - 2 x y \cos \theta_{\rm h}}, 
\label{Dbar}
\end{equation}
where $\delta^{\xi}_{\eta}$ is $1$ for $\xi$ = $\eta$ otherwise $0$. 
If we diagonalize ${\bar {\cal D}}(x,y)^{\xi}_{\eta}$ with respect to the indices $\xi$ and $\eta$, we necessarily find two eigenvalues 
of equal magnitude and opposite sign.  
For $N_{\rm h}$ = $8$ as in Fig.\ref{fig:3}, 
the hot spots are connected by the operations of $C_{4v}$ which is the symmetry of the underlying lattice. 
For a special case of $N_{\rm h}$ = $4$, e.g. the circular Fermi line with the Fermi radius $|{\bf Q}|/2$, 
the hot spots are connected by the operations of $C_{4}$ the number of which is $4$. 
The number of degeneracies is then half of the number of the symmetry operations in the group of the hot spots.

Eqs.(\ref{eigenQCP}) and (\ref{Dbar}) suggest that for $\theta_h \rightarrow 0$, all ${\bar {\cal D}}_l(\kappa) \rightarrow 0$ as 
$\kappa \rightarrow 0$ in 2d. We have explicitly found that the numerical results are consistent with this conjecture. 
In this case, as we explain below, the problem eventually acquires infinite degeneracy.

The situation is actually simpler in 3d, where it follows from Eq.(\ref{eigen3}) that $|{\cal D}_l(\kappa)|$ are less singular than  $\to$ $|\ln \kappa|$ 
as $\kappa$ $\to$ $0$ for any $l$, 
while $\epsilon$ is $1/2$, 
so that $|{\bar {\cal D}}_l(\kappa)|$ = $|{\cal D}_l(\kappa)|/\sqrt{\sum_l d_l {\cal D}_l^2}$ $\sim$ 
$\kappa^{1/2}|\ln \kappa|$, i.e.,  
${\bar {\cal D}}_l(\kappa)$ are zero for all $l$ in the limit $\kappa \to 0$.

With this knowledge of  ${\bar {\cal D}}_l(\kappa)$, we return to the RG Eqs.(\ref{beta}) to study the fixed points. 
It is straightforward to prove that ${\bar {\cal D}}_l(\kappa)$ approach constants as $\kappa \rightarrow 0$ 
(${\rm d} {\bar {\cal D}}_l / {\rm d} {\bar t}$ $\to$ 0 as ${\bar t}$ $\to$ $\infty$), as may also be seen in Fig.\ref{fig:3}. 
Therefore, it is sufficient to  analyze with $={\bar {\cal D}}_l(0)$.  
There are two possibilities for the fixed points of Eq.(\ref{beta2}). (i) ${\bar g}$ $\equiv$ ${\bar g}^* =0$. 
Then, it follows from Eqs.(\ref{beta}) that the usual multichannel problem is realized.
Due to the small channel anisotropy this is always unstable towards the single channel Fermi-liquid fixed point. 
On the other hand, a new class of singular or non-Fermi liquid (NFL) fixed points is obtained for (ii) $\sum_l d_l N_0^2f_l^{*2} = \epsilon$. 
Then the fixed-point values of  ${\bar g}^*$, ${f_l^*}$ are  solutions of the following equations;
\begin{subequations}
\label{eq:20}
\begin{equation}
\sum_{l} d_l \left( 1 + \sigma_l^{\phantom *} \sqrt{1 - 8 {\bar {\cal D}}_l(0) N_0 {\bar g}^* / \epsilon } \right)
= 8/\epsilon\;,
\label{eq:20a}
\end{equation}

\begin{equation}
\label{eq:20b}
N_0 {f}_l^* = 
\frac{\epsilon}{4} \left( 1 + \sigma_l^{\phantom *} \sqrt{1 - 8{\bar {\cal D}}_l(0) N_0 {\bar g}^*/\epsilon } \right)
\;.
\end{equation}
\end{subequations}
The fixed-point values of $j_l$ denoted by $j_l^*$ are  given by 
$j_l^*$ $=$ $f_l^* - {\bar g}^* {\bar {\cal D}}_l^{\phantom *}(0)$.  
In Eqs.(\ref{eq:20}), $\sigma_l$ is either of $\pm 1$ for each $l$. 

Of these NFL fixed points, one in which all of $\sigma_l$ are $- 1$ can be shown to be linearly {\it stable}.  
For $d$ = $2$, this stable fixed point exists only when the Fermi line is almost 
tangent at the hot spots to the boundary of the magnetic Brillouin zone, i.e. $\theta_{\rm h}\rightarrow 0$ 
(otherwise there is no solution of Eqs.(\ref{eq:20}) with $\sigma_l = -1$ for all $l$).  
For $d =3$, this fixed-point solution is always found.

When $\theta_{\rm h}\rightarrow 0$ for $d = 2$, and $d = 3$, 
a study of the solution of Eqs.(\ref{beta}) reveals that the RG flow of ${\bar g}$ and $j_l$ is 
toward the single channel Fermi-liquid fixed point (i) only 
for large initial coupling $N_0J$ $>>$ $1$, i.e. for initial $g$ $<<$ $j_l$; 
the Kondo temperature is so high that AFM correlations do not determine the fixed point. 
For the interesting weak-coupling case $N_0J$ $<<$ $1$, i.e. for initial $g$ $>>$ $j_l$,  
the RG flow is sucked into the stable one of the new class of NFL fixed points (ii),  
i.e., the fixed point of the degenerate multichannel Kondo problem with a finite ${\bar g}^*$,  
as described below, is realized at the QCP.      

For $\theta_{\rm h}$ = $0.05$ with $8$ hot spots in 2d, an explicit calculation of the stable fixed point shows that 
$|N_0{\bar g}^*|$ = $0.9205$ and that $N_0j_l^*$ = $0.06731$, $0.03895$, $0.02927$, $\ldots$, 
in order of size.  
At each value there are exactly four degenerate states in response to the enhanced degeneracy of ${\bar {\cal D}}_l(0)$. 
Thus a multiple degenerate four-channel fixed point with a finite ${\bar g}^*$ is realized 
for both signs of coupling to the localized spin. 

In the limit of $\theta_{\rm h}$ $\to$ $0$, in 2d, i.e. the Fermi surface tangent to the magnetic Brillouin zone, 
$max[|{\bar {\cal D}}_l(0)|]$ $\to$ $0$ with 
$\sum_l d_l {\bar {\cal D}}_l(0)^2$ = $1$. 
In this case, expansion of Eqs.(\ref{eq:20}) with respect to ${\bar {\cal D}}_l(0)$ 
leads us to  
\begin{equation}
N_0^2{\bar g}^{*2} = \epsilon,  \quad N_0 j_l^* = 0 \quad \mbox{for all $l$}. 
\label{eq:fixpt}
\end{equation}
So, $N_0 f_l^* = \sqrt{\epsilon}{\bar {\cal D}}_l^{\phantom *}(0) \to 0$ with 
$\sum_l d_l^{\phantom *} N_0^2  f_l^{*2}$ = $\epsilon$. 

As shown above, for $d$ = $3$,  $max[|{\bar {\cal D}}_l(\kappa)|]$ $\to$ $0$ as $\kappa$ $\to$ $0$ 
with $\sum_l d_l {\bar {\cal D}}_l(\kappa)^2$ = $1$. Then 
the exotic stable fixed point given by Eq.(\ref {eq:fixpt}) always exists.
So, the fixed point looks like the multichannel fixed point at which the number of channels is infinity. 

It is important to note that channel anisotropy is irrelevant at these stable fixed points for $d=2$ as well as $d=3$. 
This can be proved by noting that $\sum_l d_l^{\phantom *} N_0^2  f_l^{*2}$ = $\epsilon$ and 
${\rm d} {\bar {\cal D}}_l / {\rm d} {\bar t}$ $\to$ 0 as ${\bar t}$ $\to$ $\infty$. 
Then Eqs.(\ref{beta}) leads to 
${\rm d}|f_l - f_{l'}| / {\rm d}{\bar t} = |f_l - f_{l'}| (2 N_0f_l^* - \epsilon / 2 ) < 0$ 
as ${\bar t}$ $\to$ $\infty$ when ${\bar{\cal D}}_{\l}(0)$ = ${\bar{\cal D}}_{\l'}(0)$ for $l \neq l'$.  

The other fixed points given by Eqs.(\ref{eq:20}) are unstable. 
In the case of $max[|{\bar {\cal D}}_l(0)|]$ $\to$ $0$, 
$N_0^2{\bar g}^{*2}$ $=$ $\epsilon ( 1 - \epsilon n_{+}/4 )$, 
$N_0 j_l^*$ $=$  $\epsilon / 2$ for $\sigma_l$ = $1$, 
$N_0 j_l^*$ $=$  $0$ for $\sigma_l$ = $-1$, 
where $n_{+}$ is the number of channels for which $\sigma_l$ = $1$. 
Since $( 1 - \epsilon n_{+}/4 )$ must be positive or zero, there exist four (eight) unstable fixed points in two (three) dimensions 
where $\epsilon$ = $1$ ($\epsilon$ = $1/2$). 
If $\epsilon$ is assumed to be small, 
these may be related to the  unstable fixed points of a multi-channel version of the Bose-Fermi Kondo  
model in the $\epsilon$-expansion \cite{Sengupta00}. 

The difference of the results from those for the Bose-Fermi Kondo model \cite{Sengupta00, Si01} are instructive. 
In the present work, the bosons or spin-fluctuations  enter the theory only as intermediate states and not in external vertices, 
see Figs.\ref{fig:1} and \ref{fig:2}. 
This is the consistent formulation of the problem because the fluctuations arise in the first place due to the electron-electron interaction 
vertex $\lambda$. In Refs.\cite{Si01, Sengupta00}, 
$\lambda$ is implicitly included in part of the problem in defining the fluctuations but neglected in the other part, 
the conversion of the fluctuations back to fermions. 
 
The correlation functions near the fixed point for $W > W_1$ as well as the detailed properties when approaching the fixed point from $W < W_1$ 
will be presented in a longer paper. 
At this point we can only say that Eq.(\ref{eq:fixpt}): $f_{\l}^*\rightarrow 0$ suggests a decoupling of the local moment from 
the conduction electrons, while a finite $\bar {g}^*$ at the fixed point suggests that the moment responds to the AFM correlations of the host. 
This is what may be expected if the Kondo effect is deconstructed such that the local moment is at least partially recovered and the recovered moment 
participates in the AFM correlations \cite{GI05}. The infinite degeneracy at the fixed point suggests that the ground state has finite entropy. 
This degeneracy may also be understood as the prelude to the participation of the moment in the infinite range spin-wave correlations 
below the AFM transition. This infinite channel fixed point may be thought of as the analog for staggered magnetization correlations of 
what happens for growing FM correlations \cite{LM72}, where a droplet of size $\kappa^{-1}$ around the impurity leads to number of channels 
$\propto \kappa^{-(d-1)}$ \cite{MMS01}. 
However this analogy is only suggestive both because of the special condition of "hot-spots" or "hot-lines" required to get such a fixed point 
as well as the very special second property, on $f_{\l}^{*2}$, noted after Eq.(\ref{eq:fixpt}) at the fixed point.

HM and CMV are grateful to D. MacLaughlin and S. Yotsuhashi for clarifying comments. HM would also like to acknowlede T. Kato, J. Kishine, H. Kusunose, Y. Matsuda, and Y. Takada for useful discussions 
as well as the University of California, Riverside where work on this letter was partially done. CMV wishes to thank the Humboldt foundation and the condensed matter physicists at  University of Karlsruhe
for their hospitality.

\end{document}